\documentclass[a4paper,fleqn,usenatbib]{mnras}

\usepackage[T1]{fontenc}
\usepackage{ae,aecompl}

\usepackage{graphicx}	
\usepackage{amsmath}	
\usepackage{amssymb}	
\usepackage{gensymb}
\usepackage[normalem]{ulem}

\title[Spatial clustering and HOD modeling of local AGN]{Spatial clustering
  and halo occupation distribution modeling of local
  AGN via cross-correlation measurements with 2MASS galaxies}

\author[M. Krumpe et al.]{
Mirko Krumpe,$^{1}$\thanks{E-mail: mkrumpe@aip.de}
Takamitsu Miyaji,$^{2,3}$, Alison L. Coil$^{4}$ and
Hector Aceves$^{2,3}$
\\
$^{1}$Leibniz-Institut f\"ur Astrophysik Potsdam (AIP), An der Sternwarte 16, 
     14482 Potsdam, Germany\\
$^{2}$IAUNAM-E (Instituto de Astronom\'ia de la Universidad Nacional Aut\'onoma de
M\'exico Ensenada), Ensenada, Apdo. Postal 106, \\ Ensenada BC, 22800 Mexico\\     
$^{3}${\it mailing address:} IAUNAM-E (Instituto de Astronom\'ia de la Universidad Nacional Aut\'onoma de M\'exico Ensenada), \\P.O. Box 439027, San Diego, CA 92143-9027, USA\\
$^{4}$University of California, San Diego, Center for Astrophysics and
                 Space Sciences, 9500 Gilman Drive, La Jolla, CA 92093-0424, USA
}

\date{Accepted 2017 October 13. Received 2017 October 12; in original form 2017 August 17}

\pubyear{2018}

\begin{document}
\label{firstpage}
\pagerange{\pageref{firstpage}--\pageref{lastpage}}
\maketitle

\begin{abstract}
We present the clustering properties and halo occupation distribution (HOD) 
modelling of very low redshift, hard X-ray-detected active galactic nuclei (AGN) 
using cross-correlation function measurements with Two-Micron All Sky Survey   
galaxies. Spanning a redshift range of $0.007 < z < 0.037$, with a
median $z=0.024$, we present a precise AGN clustering study of 
the most local AGN in the Universe. The AGN sample is drawn from the
 {\it SWIFT}/BAT 70-month and {\it INTEGRAL}/IBIS eight year 
all-sky X-ray surveys and contains both type I and type II AGN. 
We find a large-scale bias for the full AGN sample of
$b=1.04^{+0.10}_{-0.11}$, which corresponds to a typical host dark
matter halo mass of $M_{\rm h}^{\rm typ}=12.84^{+0.22}_{-0.30}\,h^{-1} M_{\odot}$. 
When split into low and high X-ray luminosity and type I and type II
AGN subsamples, 
we detect no statistically significant differences in the large-scale bias parameters. 
However, there are differences in the small-scale clustering which are reflected
in the full HOD model results.  We find that low and
high X-ray luminosity AGN, as well as type I and type II AGN, occupy
dark matter haloes differently, with 3.4$\sigma$ and 4.0$\sigma$
differences in their mean halo masses, respectively, when split by luminosity and type. 
The latter finding contradicts a simple orientation-based AGN unification model.
As a by-product of our cross-correlation approach, we also present the first HOD model of 2MASS galaxies.
\end{abstract}




\section{Introduction}
\label{sec:introduction}

Observed AGN clustering measurements as a function of AGN parameters such as redshift, luminosity, 
black hole mass, and accretion rate have received considerable interest 
in recent years (e.g. \citealt{coil_georgakakis_2009}; \citealt{shen_strauss_2009}; \citealt{mountrichas_georgakakis_2012}; \citealt{komiya_shirasaki_2013}; \citealt{koutoulidis_plionis_2013}; \citealt{zhang_wang_2013}; \citealt{krumpe_miyaji_2015}). 
These measurements not only provide access to the typical dark matter halo (DMH) mass in which AGN reside, 
but also constrain fundamental AGN physics when compared to different theoretical models of AGN triggering 
(e.g. \citealt{allevato_finoguenov_2011}). 
Instead of determining only a typical DMH mass, more recent clustering measurements are 
able to provide detailed constraints on how AGN populate DMHs 
(see review \citealt{krumpe_miyaji_2014} and work by, e.g. \citealt {li_kauffmann_2006}; \citealt{mandelbaum_li_2009}; \citealt{padmanabhan_white_2009};
\citealt{shen_hennawi_2010}; \citealt{miyaji_krumpe_2011}) 
and thus reveal insights into how AGN are triggered and place them in a 
cosmological context.

Few studies exist of AGN clustering in the local Universe, as 
large-scale clustering measurements at very low redshift are  
challenging. First, the number density of AGN 
in the local Universe is an order of magnitude lower than at the 
peak of AGN activity at $z\sim 2$ (e.g. \citealt{miyaji_hasinger_2015};
\citealt{aird_coil_2015}). Hence, there are substantially fewer AGN per unit
comoving volume at low redshift compared to higher redshifts. 
Secondly, robust large-scale clustering measurements
require surveys that span large cosmological volumes. 
At higher redshifts ($z>1$), surveys can provide this by covering only $\sim10$ 
deg$^2$ on the sky. In the local Universe, however, only surveys that cover 
most of the sky can map a large
enough comoving volume which also contains a sufficient number of AGN 
required for robust clustering measurements. 

These issues make clustering measurements of local AGN extremely difficult, 
and only two studies of AGN clustering in the local Universe exist. 
\cite{cappelluti_ajello_2010} present the auto-correlation function (ACF)
of 199 Seyfert AGN that are spectroscopically confirmed and 
detected by {\it Swift}/BAT in the 15--55 keV all-sky survey. The sample is drawn from
\cite{ajello_costamante_2009} and covers a redshift range of 
$0.001 \lesssim z \lesssim 0.15$ (with a median $z=0.045$). 
Due to the relatively low number of pair counts, \cite{cappelluti_ajello_2010}
obtain correlation functions with large uncertainties.

\cite{li_kauffmann_2006} present the other study of AGN clustering in the local Universe. 
They use approximate 90\,000 narrow-line AGN selected from SDSS data release 4. 
These optical type II AGN are extracted from the SDSS 'main galaxy sample' and have a median $z=0.10$.
Emission-line diagnostics (\citealt{kauffmann_heckman_2003}) are used to separate narrow-line AGN from 
galaxies in this study, for which the main emphasis is to compare the clustering of AGN to inactive galaxy 
samples with the same structural properties and stellar masses.

In this paper, we undertake an effort to compute the clustering properties of local AGN, 
as several recent improvements now allow more robust measurements
and substantially tighter constraints on the halo occupation parameters of 
these AGN.  The previous work of \cite{cappelluti_ajello_2010} is based on the 36-month  
{\it Swift}/BAT data. The latest available {\it Swift}/BAT all-sky hard X-ray
survey uses 70 months of data (\citealt{baumgartner_tueller_2013}) and
contains 553 spectroscopically classified Seyfert AGN.  As hard X-rays are able to
penetrate large amounts of obscuring material, a hard X-ray-selected-AGN sample 
has the great advantage of being almost unbiased with respect to the intrinsic 
absorbing column density of the central engine. 

Additionally, \cite{huchra_macri_2012} published the three-dimensional 
distribution of nearly 45\,000
galaxies in the nearby Universe based on the Two-Micron All Sky Survey (2MASS; 
\citealt{skrutskie_cutri_2006}). This sample serves as a tracer set for the
cross-correlation technique, which our team has used extensively in previous papers 
to study the clustering properties of broad-line AGN at redshifts of
$0.07 < z < 0.50$ ({\citealt{krumpe_miyaji_2010} (paper I),
  \citealt{krumpe_miyaji_2012} (paper III), \citealt{krumpe_miyaji_2015} (paper IV)).  
Using cross-correlation measurements of AGN with a large number of galaxies in
the same volume provides many more pairs (of galaxies and AGN) at all scales to use 
in the clustering measurements, compared to 
measuring only the ACF of AGN. 
In essence, the higher density of galaxies is used to more accurately trace the 
underlying cosmic web than can be achieved using only lower density AGN 
samples.  Thus, the cross-correlation method used here is crucial to creating  more robust AGN clustering measurements in the local Universe. 
This approach has been successfully used by other clustering studies as well (e.g. \citealt{li_kauffmann_2006}; \citealt{mandelbaum_li_2009}; \citealt{mountrichas_georgakakis_2012}). 

In \cite{miyaji_krumpe_2011} we present a theoretical method that allows us 
to determine the large-scale bias parameter directly from the measured
cross-correlation function (CCF) between AGN and the galaxy tracer set, using halo 
occupation distribution (HOD) modelling. In 
\cite{krumpe_miyaji_2012} we show that this approach is strongly preferred over 
determining the bias parameter from a power-law fit of $r_0$ and $\gamma$
to the correlation function. Direct HOD modelling also allows us to constrain 
not only the large-scale bias for the AGN but also the full distribution of AGN 
in DMHs as a function of halo mass.

This paper is organized as follows. In Section~2 we describe the AGN data and samples 
and the 2MASS galaxy sample. Section~3 briefly summarizes the cross-correlation 
technique used here and how the HOD modelling is used to derive the AGN clustering 
properties.  Section~4 shows the results of our clustering measurements. 
Our results are discussed in Section~5, and 
we present our conclusions in Section~6. 

Determining the clustering properties of local AGN is of high interest to the astrophysical 
community. Powell et al. (submitted to ApJ) also initiated a study to explore local AGN clustering 
using a similar data set. As we do, they also make use of the cross-correlation technique using 
{\sl Swift} BAT AGN and 2MASS galaxies, though they use a somewhat different sample selection 
criteria and methodology to analyse the correlation function. 

Throughout this paper, all distances are measured in co-moving 
coordinates and given in units of $h^{-1}$\,Mpc, where $h= H_{\rm 0}/100$\,km\,s$^{-1}$\,Mpc$^{-1}$, unless 
otherwise stated. We use a cosmology of $\Omega_{\rm m} = 0.3$, $\Omega_{\rm \Lambda} = 0.7$, and 
$\sigma_8(z=0)=0.8$, which is consistent with the {\it Wilkinson Microwave Anisotropy Probe} data release 7 
(Table~3 of \citealt{larson_dunkley_2011}). We choose this set of cosmological parameters
over more recent results from Planck (\citealt{planck_collaboration_2016}) to be consistent with those used in our papers I--IV.
Luminosities and absolute magnitudes are calculated using $h=0.7$.
We use AB magnitudes throughout the paper. All uncertainties represent 1$\sigma$ 
(68.3\%) confidence intervals unless otherwise stated.

\section{Data}

The galaxy sample used as our tracer set for the AGN-galaxy cross-correlation 
is based on the 2MASS redshift survey (\citealt{huchra_macri_2012}). 
The AGN data are drawn from the 70-month \textit{SWIFT}/BAT
all-sky hard X-ray survey (\citealt{baumgartner_tueller_2013}) and the 
\textit{INTEGRAL}/IBIS survey (\citealt{malizia_bassani_2012, malizia_landi_2016}). 
Here we provide relevant information on each of these samples.

\subsection{2MASS Galaxies}

The 2MASS (\citealt{skrutskie_cutri_2006}) 
mapped the entire sky in the $J$, $H$, and $K_S$ bands using twin 1.3-m
telescopes located at Mount Hopkins, AZ and Cerro Tololo, Chile from 1997--2001.  
2MASS contains 470 million sources, out of which 1.6 million are extended. 
\cite{jarrett_2004} present an extended source photometric catalogue of
approximately 1 million objects with $K_S \le 13.5$ mag. This catalogue
constitutes the initial selection of sources for the redshift survey. 

The primary extragalactic goal of the 2MASS was the generation of an all-sky
galaxy redshift survey, to map the large-scale structure of the local Universe. 
Thus, a spectroscopic program
was undertaken with the goal of observing all objects with 
(i) $K_S \le 11.75$ mag and detected at $H$, (ii) $E(B-V) \le 1$ mag, and (iii) 
$|b| \ge 5\degree$ for $30\degree \le l \le 330\degree$; $|b| \ge 8\degree$
otherwise. The latter is used to avoid confusion near the galactic plane and galactic 
centre, respectively. The 2MASS magnitudes are extinction corrected. 
\cite{huchra_macri_2012} present the redshifts for 43\,533 galaxies (97.6\%) 
out of 44\,599 entries that match the aforementioned criteria.  The redshifts are 
taken from various sources (e.g. Sloan Digital Sky Survey DR8; 
6dF Galaxy Survey DR3; NASA Extragalactic Database), including spectroscopic 
follow-up observations conducted between 1998--2011.  

The galaxies within the 2MASS redshift survey span all 
morphological types (elliptical, spiral, irregular, unbarred, barred, etc.). 
Spirals dominate the sample at $z<0.03$, while ellipticals dominate 
at higher redshifts. 
Unfortunately, morphological classifications exists for only 20\,860
galaxies with $K_S \le 11.25$ or fainter if the morphological types were
available from the literature. Thus, for the entire 2MASS galaxy redshift sample 
it is not possible to select a galaxy subsample based uniformly on 
morphological classification.

\subsubsection{2MASS Galaxy Sample}
\label{2MASS_galaxy_sample}

For our 2MASS galaxy sample, we select all 2MASS redshift survey galaxies 
with $|b| \ge 8\degree$ and within a redshift range of $0.007\le z\le 0.037$. 
We compute absolute $K_S$ magnitudes ($M_{\rm K_s}$) based on the total 
extinction-corrected magnitude. Since galaxy
clustering is known to depend on luminosity (e.g.
\citealt{zehavi_zheng_2011}), 
we create a volume-limited sample by selecting only galaxies with 
$-24.4 \le M_{\rm K_s} \le -25.9$ (Fig.~\ref{2MASS_mag_z}). 
This selection ensures that 
the clustering of the galaxies used in this paper as the tracer set 
remains constant over the redshift range used here.  This absolute magnitude
limit results in a tracer sample that is restricted to $z \le 0.037$.

In the process of performing the analyses for this paper, we initially used 
multiple redshift ranges. 
We initially split the galaxy sample into low- and high-redshift subsamples and 
computed the ACFs of these subsamples. We noticed 
that including galaxies with redshifts above $\sim$0.04 leads to increasingly 
different clustering properties between the low and high-redshift subsamples. 
This is due to the ratio of spirals to 
ellipticals dropping substantially above $z\sim0.04$, reflecting the increased 
fraction of elliptical galaxies leading to a higher clustering signal. Because 
morphological types are only available for approximately half of the galaxy sample, it is
not possible to create subsamples that contain only spiral or elliptical
galaxies and hence overcome this issue. 

The final redshift range used here is a trade-off between (i) maximizing the number 
of possible cross-correlated pairs between 2MASS galaxies and AGN, and 
(ii) identifying a redshift range in which the 2MASS galaxies show little to no 
evolution of the large-scale bias parameter as a function of redshift. 
At $z \lesssim 0.037$ the fraction of spiral and elliptical galaxies is roughly constant
(maximal derivation from the median value $\pm$8\%),
which results in very similar clustering amplitudes
in the low and high-redshift subsamples (for details see Section~\ref{compare_gal_lowhighz}).
This is crucial for our study, as here we use the CCF with 
the tracer sample across the full redshift range of $0.007 \le z \le 0.037$ to 
decrease the error bars, and this approach relies on the galaxy tracer set 
clustering properties not evolving strongly over the full redshift range used. 
We therefore strongly recommend that the 2MASS galaxy sample above $z \sim 0.4$ 
not be used for cross-correlation studies, unless one is willing to accept 
significantly increased systematic errors.  

The depth of 2MASS depends on 
variations in, for example, the seeing and sky background.
The magnitude limits are therefore selected such that the
source sensitivity and detection are complete under all observing conditions 
for the survey. Consequently, no spatial window function is required. 
Within and around the Magellanic clouds, we recognize no significant excess 
or deficit in number density of 2MASS sources with $K_S \le 11.75$ mag. 
Thus, areas covered by the Magellanic clouds are included.
The final sample contains 10\,900 objects covering 35\,512
deg$^2$ of the sky. The number density of the volume-limited 
sample used here is $n = (2.29\pm 0.13) \times 10^{-3}$ 
$h^3$ Mpc$^{-3}$. 

To estimate the uncertainties in the number density and 
the clustering amplitude (see Section~\ref{errorAnalysis}), we divide the
survey area into 22 subsections of $\sim$1500--1850 deg$^2$ each. 
The decision to create relatively large subsections and therefore a moderate number 
of jackknife samples is based on the largest scales that we aim to probe 
with our clustering measurements. 
The final 2MASS sample covers a redshift range of $0.007 \le z \le 0.037$, 
with a median redshift of $\langle z \rangle=0.029$. We measure the correlation 
function to a projected distance of  30 $h^{-1}$ Mpc. The angular 
distance of 30 $h^{-1}$ Mpc at a  
redshift of $z=0.01$ corresponds to $\sim$40$^{\circ}$. Thus to include this 
scale in each jackknife subsection, $\sim$1600 deg$^2$ are required for 
each subsection.

\begin{figure}
\resizebox{\hsize}{!}{
 \includegraphics[bbllx=77,bblly=390,bburx=550,bbury=700,angle=0]{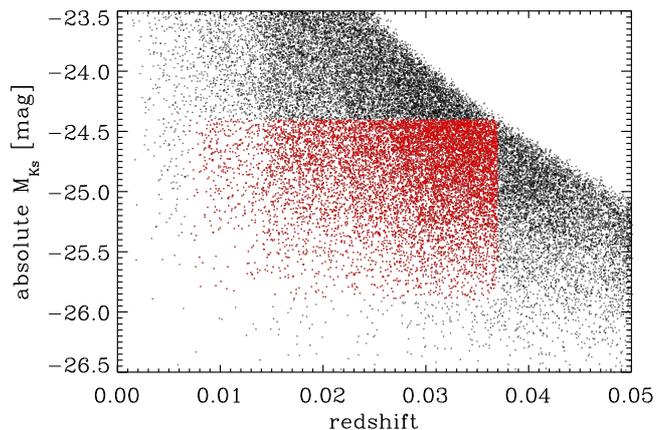}}
 \caption{Absolute $K_S$ magnitude versus redshift for galaxies in the 2MASS redshift
   survey (black). Galaxies in red indicate our tracer sample of
   10\,900 galaxies used for cross-correlation measurements with $0.007 \le z \le 0.037$ and $-24.4 \le M_{K_S} \le -25.9$.}
 \label{2MASS_mag_z}
\end{figure}

\subsubsection{2MASS Random Sample}
In order to compute the correlation function, a random galaxy sample is required (see
Section~\ref{clusteringMeasurements} below). Since no spatial window function is
needed for the observed 2MASS galaxy sample, we randomly distribute objects 
on the sky. We select only those sources with $|b| \ge 8\degree$ to match the
observed sample. The random catalogue contains 500 times the number of 
2MASS galaxies in our tracer set. This ensures that we minimize the
uncertainties introduced due to statistical effects of the random catalogue, even
for pairs at the smallest scales measured here. 

The corresponding redshift for a random object is randomly chosen from the 
smooth redshift distribution of the 2MASS galaxy sample tracer set 
(Fig.~\ref{2MASS_z_dist}).
The smoothing applies a least-squares (\citealt{savitzky_golay_1964})
low-pass filter to the observed 2MASS redshift distribution.

\begin{figure}
\resizebox{\hsize}{!}{
 \includegraphics[bbllx=77,bblly=390,bburx=538,bbury=700,angle=0]{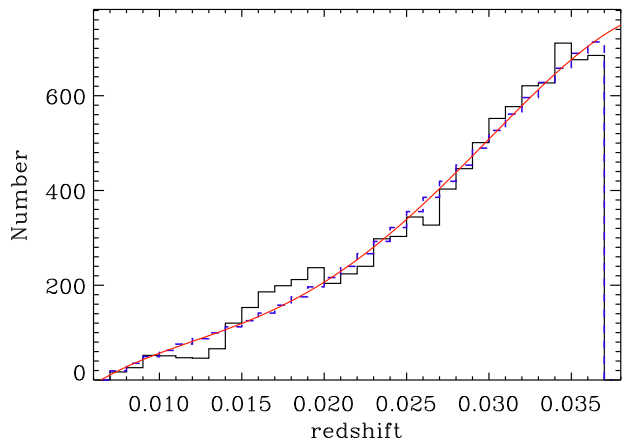}}
 \caption{Redshift histogram for the observed 2MASS galaxy sample used as our tracer set
   (black histogram), the smoothed redshift profile (smooth red solid line) and the random sample (blue-dashed, renormalized to the distribution of 2MASS galaxies).}
 \label{2MASS_z_dist}
\end{figure}

\subsection{\textit{SWIFT}/BAT AGN}
The {\it SWIFT} gamma-ray burst observatory
(\citealt{gehrels_chincarini_2004}) has continuously observed the hard X-ray sky
with the Burst Alert Telescope (BAT) in an energy range of 14--195 keV since 
its launch in 2004. The BAT is a coded-mask telescope with a field of view of 
$\sim$60$^{\circ} \times 100^{\circ}$. The effectively random pointing strategy 
developed to detect and
study gamma-ray bursts is also well suited for performing an all-sky survey
with uniform sky coverage. Due to its large field of view, BAT covers 
approximately one-sixth of the sky with each pointing. 

As the total observation time accumulates  
higher sensitivity is reached, which combined with an improved data reduction pipeline 
results in more detected sources. In eight-band mosaics, sources are determined by
identifying 4.8$\sigma$ events, compared to their surrounding
neighbours. The sources are then cross-correlated with archival X-ray images
from high-resolution instruments and extended objects in the 2MASS
all-sky survey. The flux of known bright sources, which are detected with a significance greater
than 6$\sigma$, are subtracted and the source detection is performed on the resulting 
cleaned images. This survey is the most sensitive hard X-ray all-sky survey in existence 
and reaches a flux of 
$1.3 \times 10^{-11}$ erg s$^{-1}$ cm$^{-2}$ over 90\% of the sky (14--195 keV band). 
A detailed description on the 70-month \textit{SWIFT}/BAT
all-sky hard X-ray survey can be found in 
\cite{baumgartner_tueller_2013}. 

NED and SIMBAD are used to determine the source type and redshift of the 
counterparts. Almost 60\% of the 70-month 
\textit{SWIFT}/BAT all-sky hard X-ray survey classifications are AGN. 
\cite{baumgartner_tueller_2013} list subtypes of AGN such as Seyfert I (Sy1.0--1.5), 
Seyfert II (Sy1.7--2.0) and QSOs (luminosities larger than 10$^{45}$ erg s$^{-1}$). 
For all these sources the 14--195 keV luminosities as well as their photon indices 
(based on a fit to the eight-channel spectra) are provided.

\subsection{\textit{INTEGRAL}/IBIS AGN}
The Imager on Board the \textit{INTEGRAL} Satellite (IBIS) also provides an all-sky hard X-ray source catalogue. This coded mask instrument with an energy range of 15 keV -- 10 MeV features higher angular resolution ($\sim$12 arcmin, FWHM) compared to the $\sim$20 arcmin resolution of \textit{SWIFT}/BAT. \cite{bird_bazzano_2016} published the \textit{INTEGRAL}/IBIS all-sky hard X-ray source catalogue of all the data publicly available starting from the 12th orbit of \textit{INTEGRAL} (end of 2002) up to the end of 2010.

\cite{malizia_bassani_2012, malizia_landi_2016} present several hundred AGN 
detected by \textit{INTEGRAL}, for which they also provide the optical classification, 
redshift, X-ray column density measurements and 20--100 keV luminosities.
The counterpart classification uses literature and NED searches, as well as 
their own follow-up spectroscopy. 
The AGN are further distinguished into Sy1, Sy1.2, Sy 1.5, Sy1.9, Sy2 and QSO. 
In their 2016 paper they list additional objects since 2010 and only for those they also 
provide the photon index.
The primary goal of this survey is to study the absorption properties in a 
large sample of hard X-ray selected, and thus unbiased, AGN sample. For 
more details on the \textit{INTEGRAL}/IBIS AGN catalogues see \cite{malizia_bassani_2012, malizia_landi_2016}.

\subsection{Combined \textit{SWIFT/INTEGRAL} AGN Sample}
\label{combined_AGN_sample}

In order to maximize the number of hard X-ray-selected AGN to use for our 
clustering measurements, we combine the 70-month \textit{SWIFT}/BAT AGN 
and the \textit{INTEGRAL}/IBIS AGN catalogues. This is possible as both 
samples cover a similar parameter space in redshift, luminosity, column density, and 
photon index.  

Only AGN within a redshift range of $0.007 \le z \le 0.037$, matching our 
2MASS galaxy redshift range, are considered. Furthermore, an object is only included if 
$|b| \ge 8\degree$.
We begin with the \textit{SWIFT}/BAT AGN 
sample, which in this redshift range contains 112 type I and 141 type II AGN. We then 
identify additional \textit{INTEGRAL} AGN from the 2012 and 2016 catalogues that are 
not included in the 70-month \textit{SWIFT}/BAT AGN sample. 
Identical to the \textit{SWIFT}/BAT source classification, \textit{INTEGRAL} 
subclassifications of Sy1.0--1.5 and Sy1.9--2.0 are classified as 
type I and type II AGN, respectively. QSOs are included in the type I samples 
for both \textit{SWIFT} and \textit{INTEGRAL}. 
The 2012 \textit{INTEGRAL} AGN catalogue adds one additional type I and eight type II AGN to our AGN sample. The 2016 \textit{INTEGRAL} AGN catalogue adds an additional eight 
unique type II AGN. Thus, the final sample contains 274 AGN (including four \textit{SWIFT} AGN with the classification ``other AGN'', but excluding blazar/BL Lac objects) of which 113 and 157 are type I and type II AGN, respectively. 

Figs.~\ref{AGN_redshift}--\ref{AGN_LX} show the 
redshift and luminosity distributions of the full AGN sample, divided into type I 
and type II subsamples. Although subject to low number statistics, the full, type I, 
and type II AGN samples have similar redshift and luminosity distributions. 
We convert the 20--100 keV \textit{INTEGRAL} luminosities into the 14--195 keV \textit{SWIFT} energy band by multiplying the 20--100 keV \textit{INTEGRAL} luminosity with a factor of 1.65. This value was determined from 
WebPIMMS{\footnote{\tt{https://heasarc.gsfc.nasa.gov/cgi-bin/Tools/w3pimms/w3pimms.pl\\Based on work by Koji Mukai, Michael Arida and Ed Sabol}}}, using a median \textit{SWIFT} AGN photon index of 1.93. 

We check whether the same AGN received identical \textit{SWIFT} and \textit{INTEGRAL} type classifications. Out of 82 AGN in common between the \textit{SWIFT} AGN sample 
and the \textit{INTEGRAL} 2012 AGN sample, 78 agree in their AGN classifications. 
Additionally, Fig.~\ref{NH_type} shows that there is no degeneracy between X-ray luminosity and column density for the AGN in our sample.

Depending on its sky position, we assign each AGN to a corresponding jackknife 
subsection. This procedure is identical to the one used for the 2MASS galaxy sample (see Section~\ref{2MASS_galaxy_sample}). The selected 2MASS galaxy sample and the AGN sample have 113 objects in common. The distribution of the finally selected 2MASS galaxy sample and hard X-ray-selected AGN are shown in Fig.~\ref{AGN_2MASS_skyposition}.

\begin{figure}
\resizebox{\hsize}{!}{
 \includegraphics[bbllx=93,bblly=374,bburx=552,bbury=695,angle=0]{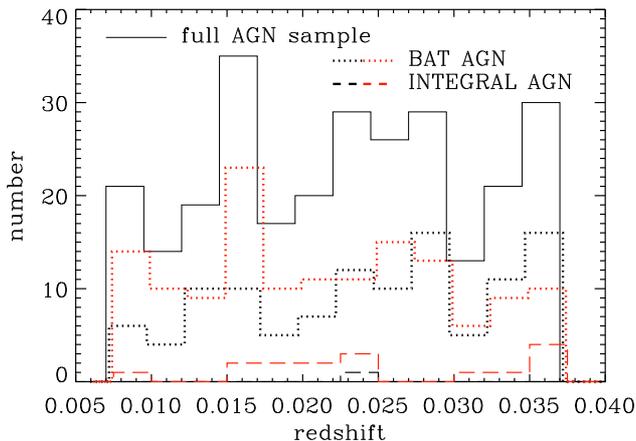}}
 \caption{Redshift histogram of the full AGN sample (black solid line), divided into subsamples based on the catalogue used for identification (dotted line -- \textit{SWIFT}/BAT, dashed line -- \textit{INTEGRAL}/IBIS) and AGN type classification (black -- type I AGN, red -- type II AGN). For clarity, we slightly shift the redshift values of the BAT and \textit{INTEGRAL} subsamples in the histogram.}
 \label{AGN_redshift}
\end{figure}

\begin{figure}
\resizebox{\hsize}{!}{
 \includegraphics[bbllx=75,bblly=374,bburx=534,bbury=695,angle=0]{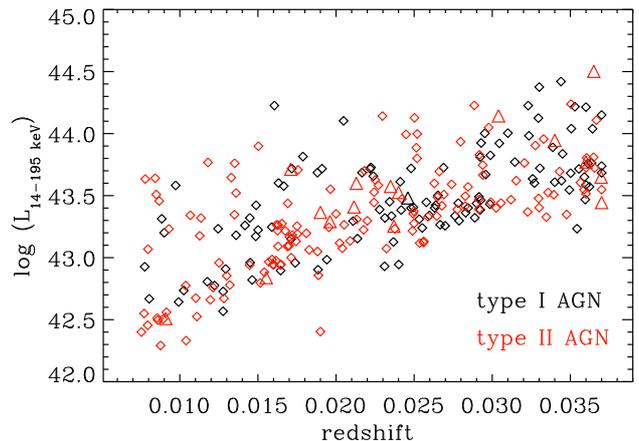}}
 \caption{X-ray luminosity versus redshift for the full AGN sample. The different symbols and colours represent the different catalogues used for identification and the AGN classification types: diamonds -- \textit{SWIFT}/BAT, triangles -- \textit{INTEGRAL}/IBIS, black -- type I AGN and red -- type II AGN.}
 \label{AGN_LX_z}
\end{figure}

\begin{figure}
\resizebox{\hsize}{!}{
 \includegraphics[bbllx=80,bblly=374,bburx=550,bbury=695,angle=0]{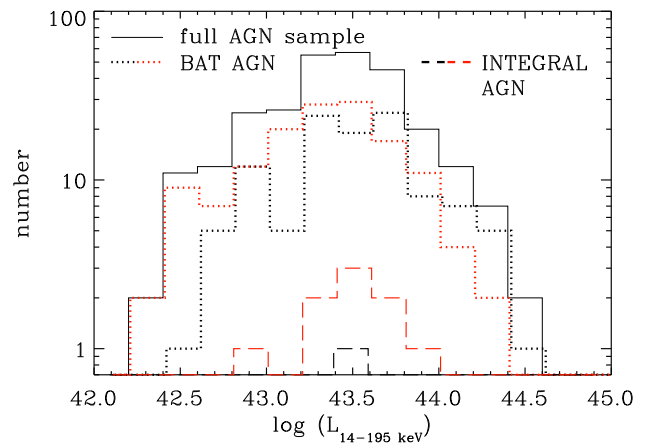}}
 \caption{X-ray luminosity histogram of the full AGN sample (black solid line) and subsamples based on identification catalogue and AGN classification type. Line types and colours are identical to Fig.~\ref{AGN_redshift}.}
 \label{AGN_LX}
\end{figure}

\begin{figure}
\resizebox{\hsize}{!}{
 \includegraphics[bbllx=81,bblly=374,bburx=534,bbury=695,angle=0]{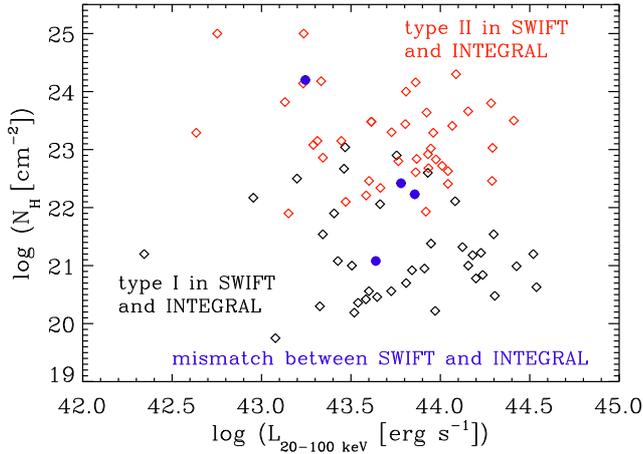}}
 \caption{X-ray luminosity in the 20--100 keV band versus intrinsic column density $N_{\rm H}$ for AGN that fulfil our selection criteria and are found in both the \textit{SWIFT} and \textit{INTEGRAL} 2012 catalogues. Black (type I) and red (type II) open symbols represent those sources that have the same AGN subclass in each catalogue. Filled blue circles show sources that have different classifications in the two catalogues.}
 \label{NH_type}
\end{figure}

\begin{figure}
\resizebox{\hsize}{!}{
 \includegraphics[bbllx=81,bblly=374,bburx=534,bbury=695,angle=0]{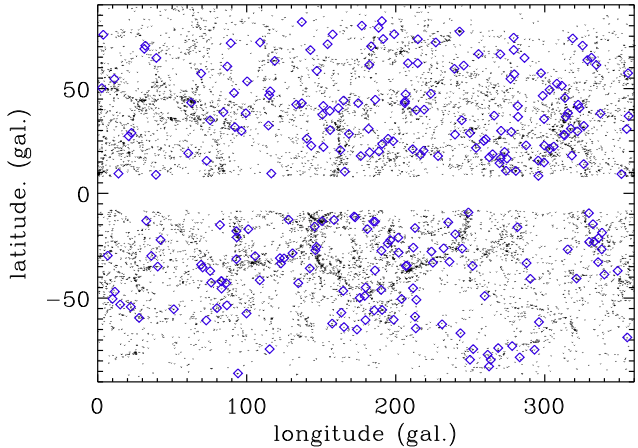}}
 \caption{ Distribution in galactic coordinates of the 10\,900 2MASS galaxies in the tracer set (black dots) and 274 AGN selected using \textit{SWIFT} and \textit{INTEGRAL} catalogues (blue diamonds). Note the restriction of $|b| \ge 8\degree$ for both, the 2MASS galaxy sample and the AGN sample.}
 \label{AGN_2MASS_skyposition}
\end{figure}

\begin{figure}
\resizebox{\hsize}{!}{
 \includegraphics[bbllx=91,bblly=368,bburx=546,bbury=695,angle=0]{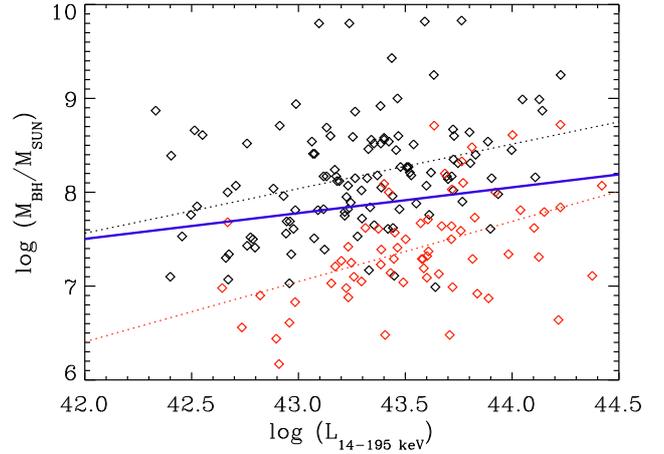}}
 \caption{Black hole mass estimates from Koss et al.~(2017)  
 versus X-ray luminosity for AGN in the 70-month \textit{SWIFT}/BAT catalogue. 
Black data points show mass estimates derived from stellar velocity dispersion measurements, while red data points show estimates derived from single epoch spectroscopy. The corresponding  $\chi^2$ minimization linear fits are shown as black and red dotted lines. The linear fit of the combined sample is shown as a blue solid line.}
 \label{Mbh_LX_koss}
\end{figure}

\subsection{Black hole mass estimates}
\label{black_hole_mass_estimates}
\cite{koss_trakhtenbrot_2017} present black hole mass estimates for a subset of AGN detected
in the \textit{SWIFT}/BAT 70-Month catalogue. For 201 objects (31\% of the complete \textit{SWIFT}/BAT AGN sample), these estimates are derived from stellar velocity dispersion measurements.
For 227 objects the estimates are single-epoch virial black hole masses derived from broad emission lines. In total 415 unique objects have black hole mass estimates from one or both of these techniques.  Applying our redshift cuts and restricting to $|b| \ge 8\degree$ for the complete \textit{SWIFT}/BAT AGN sample, 181 objects out of 274 in total have black hole mass estimates (119 from stellar velocity dispersion measurements, 69 from single epoch spectroscopy and 7 from both methods). We remove 11 sources that have disagreement between BASS and the 70-month \textit{SWIFT}/BAT AGN catalogue. According to the latter these sources are not classified as AGN.

In our redshift range, Fig.~\ref{Mbh_LX_koss} demonstrates that black hole masses based on stellar velocity dispersion have, on average, substantially higher masses than estimates from single-epoch spectra. For the seven objects that have mass estimates from both methods, the masses derived using stellar velocity dispersion are on average 0.8 dex higher than those based on single-epoch spectroscopy.

\subsection{AGN subsamples}

In addition to the full AGN sample, which contains 274 sources, we also create 
subsamples according to different available AGN parameters (see Table~\ref{overview}). 

First, we split the full AGN sample at the median X-ray luminosity of log $(L_{\rm 14-195\,keV}/$[erg s$^{-1}])= 43.42$.  The median log $L_{\rm X}$ between the low- and high-$L_{\rm X}$ samples differs by 0.5 dex. Comparing the sources for which a photon index is available, the median photon indices of both samples are very similar. The fractions of type I and type II objects in each sample are slightly different (low-$L_{\rm X}$ AGN: 36\% type I, 64\% type II; high-$L_{\rm X}$ AGN: 47\% type I, 53\% type II).

Secondly, we select the 170 objects classified as AGN in \cite{baumgartner_tueller_2013} for which black hole mass estimates are available from \cite{koss_trakhtenbrot_2017}. We split the resulting sample at the median black hole mass of log ($M_{\rm BH}/M_\odot)=7.88$. The low and high black hole mass samples contain 86 and 84 objects, respectively. The median black hole masses and redshifts of these samples are log ($M_{\rm BH,low}/M_\odot)=7.39$ and log ($M_{\rm BH,high}/M_\odot)=8.39$ and $\langle z_{\rm low} \rangle =0.022$ and $\langle z_{\rm high} \rangle =0.026$. The low-$M_{\rm BH}$ sample contains 50 type I and 35 type II AGN, with the remaining three  classified as 'other AGN'. The high-$M_{\rm BH}$ sample contains 27 type I and 56 type II AGN, with one 'other' AGN.

Thirdly, we split into type I and type II AGN classification (113 sources are in the type I sample, 157 sources are in the type II sample). 
The samples have median X-ray luminosities of 
log $(L_{\rm 14-195\,keV}/$[erg s$^{-1}])=43.45$ and 43.36 for the type I and type II samples, respectively, with corresponding photon indices of 2.01 and 1.85.
Fig.~\ref{NH_type} illustrates that a split into type I and type II subsamples is very similar to a split in intrinsic column density $N_{\rm H}$. Since we do not have $N_{\rm H}$ for all sources, we consider the results splitting into type I and type II as an estimate for 
the clustering dependence on $N_{\rm H}$ below.

Fourthly, we split the sample with respect to a photon index of $\Gamma=1.93$. For these subsamples, we use only the \textit{SWIFT} and \textit{INTEGRAL} 2016 AGN catalogues, 
as only these catalogues contain photon indices. The low- and high-$\Gamma$ samples have median photon indices of $\Gamma_{\rm low}=1.76$ and $\Gamma_{\rm high}=2.07$. 
The corresponding X-ray luminosities are very similar, but the fractions 
of type I and type II AGN in each sample are quite different (low-$\Gamma$ AGN: 28\% type I, 72\% type II; high-$\Gamma$ AGN: 60\% type I, 40\% type II).

\section{Methodology}
\subsection{Cross-correlation clustering measurements}
\label{clusteringMeasurements}

We follow the same methodology as described in Section 4 of Papers III and IV and repeat the salient features here. 
We measure the two-point correlation function $\xi(r)$ (\citealt{peebles_1980}), which measures 
the excess probability $dP$ of finding an object in a volume 
element $dV$ at a distance $r$ from another randomly chosen object. The ACF measures the spatial clustering of objects in the same sample, 
while the CCF measures the clustering of objects in 
two different samples. We determine the ACF of the 2MASS galaxy sample and the 
CCF between the AGN sample (and subsamples) with 2MASS galaxies. 

For both types of correlation functions, we use the correlation estimator of \cite{davis_peebles_1983} in the form
\begin{equation}
\label{DD_DR}
 \xi(r)= \frac{DD(r)}{DR(r)} -1\ ,
\end{equation}
where $DD(r)$ is the number of data-data pairs with a separation $r$ and $DR(r)$ 
is the number data-random pairs. Both pair counts have been normalized by the number density of
data and random points. This simple estimator has several major advantages and results in only 
a marginal loss in the S/N when compared to more advanced estimators 
(e.g., \citealt{landy_szalay_1993}). The estimator requires the generation of a random 
catalogue only for the tracer set. The selection effects of the AGN sample, including 
possible non-uniformity of the flux-limit of the catalogue, are extremely difficult to model. 
Thus, accepting a marginal loss in the signal-to-noise (S/N) leads to the greater benefit of smaller systematic
 uncertainties due to the lack of the requirement of modelling the very complex AGN selection function. 

To separate the effect of redshift distortions, the correlation 
function is measured as a function of two components of the separation vector 
between two objects, i.e., one perpendicular to ($r_p$) and the other along 
($\pi$) the line of sight. $\xi(r_p,\pi)$ is thus extracted by counting 
pairs on a two-dimensional grid of separations $r_p$ and $\pi$. 
We obtain the  projected correlation function $w_p(r_p)$ by integrating $\xi(r_p,\pi )$ along
the $\pi$ direction.

We measure $r_p$ in the range of 0.05--30 $h^{-1}$ Mpc in 14 logarithmic bins. 
We compute $\pi$ in steps of 5 $h^{-1}$ Mpc in the range $\pi=0-200$ $h^{-1}$ Mpc. 
To derive $\pi_{\rm max}$, we compute $w_p(r_p)$ for a set of $\pi_{\rm max}$ ranging 
from 10 to160 $h^{-1}$ Mpc in steps of 10 $h^{-1}$ Mpc. We then fit $w_p(r_p)$ for 
$r_p = 0.3--3$0 $h^{-1}$ Mpc, using a fixed $\gamma = 1.9$, and determine the 
correlation length $r_{\rm 0}$ for the individual $\pi_{\rm max}$ measurements 
using the following equation:
\begin{eqnarray}
\label{powerlaw}
 w_p(r_p) &=& r_p\left(\frac{r_{\rm 0}}{r_p}\right)^{\gamma}\,\frac{\Gamma(1/2)\Gamma((\gamma-1)/2)}{\Gamma(\gamma/2)}.
\end{eqnarray}
For the 2MASS galaxy ACF and the AGN--2MASS CCFs the $r_{\rm 0}$ signal has 
already saturated at $\pi_{\rm max}=40$ $h^{-1}$ Mpc. For larger values of $\pi_{\rm max}$ 
the corresponding correlation lengths
do not change by more than 1$\sigma$, and the uncertainties increase with increasing $\pi_{\rm max}$ values.

\subsection{Error Analysis}
\label{errorAnalysis}

The error analysis is also identical to Papers I, III and IV.
We use the jackknife resampling technique to estimate the measurement errors
based on the covariance matrix $M_{ij}$, which reflects the degree to which bin $i$ is 
correlated with bin $j$. 

As discussed above in Section~\ref{2MASS_galaxy_sample}, we divide the survey area 
into $N_{\rm T}=22$ subsections, determined such that each subarea spans the largest scale studied in this paper.  
The $N_{\rm T}$ jackknife-resampled correlation functions define the 
covariance matrix: 
\begin{eqnarray}
\label{jackknife}
 M_{ij} = \frac{N_{\rm T} -1}{N_{\rm T}} \left[\sum_{k=1}^{N_{\rm T}} \bigg(w_k(r_{p,i})-<w(r_{p,i})>\bigg)\right.\nonumber\\
          \times \bigg(w_k(r_{p,j})-<w(r_{p,j})>\bigg)\bigg] \,  
\end{eqnarray}
We calculate $w_p(r_p)$ $N_{\rm T}$ times, where each jackknife sample excludes one 
section, $w_k(r_{p,i})$ and $w_k(r_{p,j})$ are from the $k$th jackknife samples of the AGN ACF and $<w(r_{p,i})>$ and $<w(r_{p,j})>$ are the averages over all of the 
jackknife samples. The uncertainties represent 1$\sigma$ (68.3\%) confidence intervals ($\sigma_{i}=\sqrt{M_{ii}}$).


\subsection{HOD modelling}

We interpret our results using HOD modelling, following
an approach similar to that presented in paper II.  We use the HOD approach to obtain
linear bias parameters as well as to investigate differences in the
measured CCFs among various subsamples of the {\sl Swift} BAT/Integral IBIS AGN with 2MASS galaxies beyond 
differences in the bias parameters alone. In this paper we do not intend to present full constraints of the AGN HODs. 
Thus the HOD model of the AGN is deliberately made simple. 

The outline of our approach is as follows:
\begin{enumerate}
\item We first apply the HOD modelling technique to the ACF of the 2MASS galaxies and
  find accurate central and satellite HODs using a correlated $\chi^2$ fit with a four free parameter model.
  The number density constraint is also included.
  The best-fitting parameter search is made with a Markov chain Monte Carlo (MCMC) method.
\item In order to model the CCF between the 2MASS galaxies and AGN, two sets of HODs are needed,
  the HOD of the 2MASS galaxies derived in the previous step and that of the AGN. Since the 2MASS galaxy 
  sample size is much larger than that of those of the AGN samples,  
  and therefore the statistical significance of the 2MASS galaxy ACF is much better than that of the 2MASS galaxy--AGN 
CCF, we use the best-fitting 2MASS galaxy HOD derived above and a very simplified two free parameter HOD model for the AGN. 
\item By performing a correlated $\chi^2$ fit to the CCF, we constrain the two free parameters for the AGN HOD. HOD modelling of the 2MASS ACF and the 2MASS galaxy--AGN CCF is performed on scales of $0.05 \leq r_{\rm p} \leq 30$ $h^{-1}$ Mpc.  For the CCF AGN subsamples we impose a minimum $r_{\rm p}$ such that the number of pairs in the bins used for the fit have at least 16 pairs, in order to apply robust $\chi^2$ statistics.
\end{enumerate}

For consistency with our previous papers, we use the bias parameter as a function of halo mass relation 
of \citet{tinker_weinberg_2005} and
the halo mass function of \citet{sheth_mo_2001} in our HOD modelling. A number of improvements have been made to our HOD modelling since paper II. We use the software {\tt camb}\footnote{\url{http://camb.info/}} (by Anthony Lewis and Anthony Challinor) to generate the linear power spectrum for the cosmological parameters specified in Section~\ref{sec:introduction}. The effects of halo--halo collision and scale-dependent bias to the two-halo term described in appendix B of \citep{tinker_weinberg_2005} are fully included.

This removes the need to ignore the results in the $r_{\rm p}$ range around the transition between one-halo and two-halo dominated regimes
(see paper II). Instead of using equation (7) of paper II, where $w_{\rm p}(r_{\rm p})$ for one- and two-halo terms are transformed from the respective model power spectra through the zeroth-order Bessel function of the first kind, here we use
\begin{eqnarray}
w_{\rm p,1h}(r_{\rm p})&=& 2\int_{0}^{\pi_{\rm max}} \left[1+\xi_{\rm 1h}\left(r_{\rm p},\pi\right)\right] d\pi.\nonumber \\ 
w_{\rm p,2h}(r_{\rm p})&=& 2\int_{0}^{\pi_{\rm max}} \xi_{\rm 2h}\left(r_{\rm p},\pi\right)d\pi. 
\label{eq:wp_xi}
\end{eqnarray}

Hereafter, the quantities for AGN are represented by a subscript 'A', 
2MASS galaxies by 'G' (representing {\it galaxies}), and CCF between the two by 'AG'.  
We denote the galaxy HODs at the halo centre and of satellites by 
$\langle N_{\rm G,c}\rangle (M_{\rm h})$ and $\langle N_{\rm G,s}\rangle (M_{\rm h})$, 
respectively. Now 
$\langle N_{\rm G}\rangle (M_{\rm h})=\langle N_{\rm G,c}\rangle (M_{\rm h})+\langle N_{\rm G,s}\rangle (M_{\rm h})$. 
Likewise, the HODs of the AGN at the halo centres, of satellites, and the sum of 
the two are denoted by 
$\langle N_{\rm A,c}\rangle (M_{\rm h})$, $\langle N_{\rm A,s}\rangle (M_{\rm h})$ and 
$\langle N_{\rm A}\rangle (M_{\rm h})$.

\subsubsection{HOD of 2MASS Galaxies}
\label{sec:2masshod}

For the HOD modelling of the 2MASS galaxy sample, we use the five parameter model of \citet{zheng07}:
\begin{eqnarray}
\langle N_{\rm G,c}\rangle (M_{\rm h}) &=& 
\frac{1}{2}\left[1+\mathrm{erf}\left(\frac{\log M_{\rm h}-\log M_{\rm
min}}{\sigma_{\log M}}\right)\right]\nonumber\\
\langle N_{\rm G,s}\rangle (M_{\rm h}) &=& \langle N_{\rm G,c}\rangle (M_{\rm h})\;
\left(\frac{M_{\rm h}-M_0}{M_1^\prime}\right)^{\alpha_{\rm s}},
\label{eq:zheng07hod}
\end{eqnarray}
where the five model parameters are $M_{\rm min}$, $\sigma_{\log M}$, $M_0$, $M_1^\prime$, and
$\alpha_{\rm s}$. HOD studies on various galaxy samples by \citet{zehavi_zheng_2011} show $M_{\rm min}\approx M_0$
and $M_0$ becomes poorly constrained when it is allowed to vary freely \citep{skibba2015}. Thus,
we impose $M_{\rm min}= M_0$.

Best fitting parameters are searched for by minimizing correlated $\chi^2$:
\begin{eqnarray}
\chi^2 &=& \sum_{ij}\{[w_{\rm p}(r_{{\rm p},i})-w_{\rm p}^{\rm mdl}(r_{{\rm p},i})]M^{-1}_{ij}\times\nonumber\\
       & & [w_{\rm p}(r_{{\rm p},j})-w_{\rm p}^{\rm mdl}(r_{{\rm p},j})]\} +\nonumber\\
       & &  (n_{\rm G}-n_{\rm G}^{\rm mdl})^2/\sigma_{n_{\rm G}}^2,
\label{eq:acf_chi2}
\end{eqnarray}
where the quantities from the model are indicated by a superscript 'mdl', $M_{ij}$ is the covariance 
matrix (equation~\ref{jackknife}), $n_{\rm G}$ the number density of 2MASS galaxies and $\sigma_{n_{\rm G}}$ is its 1$\sigma$ error
respectively. We measure $n_{\rm G}=(2.29\pm 0.13)\times 10^{-3}\,h^3\,{\rm Mpc^{-3}}$, where the 1$\sigma$ error is estimated
by jackknife resampling.

The best-fitting parameter search is made using an MCMC method with the MCMC-F90 library of  
Marko Laine\footnote{\url{http://helios.fmi.fi/~lainema/mcmcf90/}}, which has been linked to our HOD model
calculation software.

\begin{figure}
\resizebox{\hsize}{!}{ \includegraphics[bbllx=69,bblly=210,bburx=514,bbury=645,angle=0]{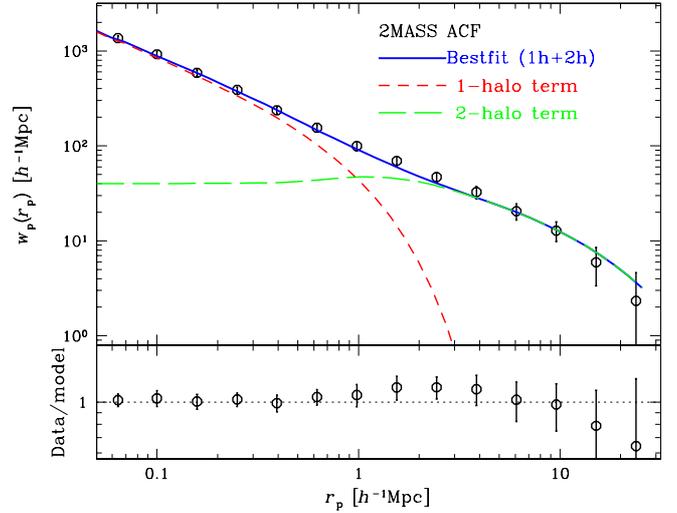}}
\caption{The 2MASS galaxy ACF (black circles) with 1$\sigma$ error bars is shown with the best-fitting HOD model (blue solid line). The one-halo and two-halo terms of the HOD model are also shown in red short dashed and green long dashed lines, respectively. The lower panel shows the fit residuals in terms of the data/model ratio.}
 \label{fig:2massacfhod}
\end{figure}

\subsubsection{HOD of {\it Swift} BAT AGN Samples}

We fit HOD models to the CCFs between the 2MASS galaxy and the AGN (sub)samples. To calculate
the expected CCF $w_{\rm p}(r_{\rm p})$, we use the best-fitting HOD of the 2MASS galaxies derived above
and a parametrized model of the AGN HOD. Because the AGN sample size is much smaller than the 2MASS sample size, 
the HOD of the 2MASS galaxies is fixed to the best-fitting values from the ACF. As discussed above, we use a very simple 
HOD model for the AGN samples, of the form
\begin{eqnarray}
\langle N_{\rm A,c}\rangle &=& f_{\rm A}\Theta(M_{\rm h}-M_{\rm min}), \nonumber \\
\langle N_{\rm A,s}\rangle &=& f_{\rm A}\Theta(M_{\rm h}-M_{\rm min})\left[(M_{\rm h}-M_0)/M_1^\prime\right])^{\alpha_{\rm s}},  
\label{eq:step_pl}
\end{eqnarray}
where $f_{\rm A}$ is the AGN fraction (duty cycle) among central galaxies at $M_{\rm h}\ga M_{\rm min}$ and $\Theta(x)$ is a step function that has the value of 0 at $x<0$ and 1 at $x\geq 0$ respectively. Like the case of the galaxy ACF, we set $M_0=M_{\rm min}$. Note that this is essentially the same model as model B of paper II except that here we
use $M_0=M_{\rm min}$ rather than zero. Furthermore, we set $M_1^\prime/M_{\rm min}=10$, which approximately represents the HODs of most luminosity-threshold nearby galaxy samples (Section~\ref{sec:2masshod}; \citealt[][]{zehavi_zheng_2011}). We do not use AGN density constraints, and the value of $f_{\rm A}$, which is the global normalization of the AGN HOD, does not affect the resulting CCF. Thus our free parameters are $\log M_{\rm min}$ and $\alpha_{\rm s}$. 

We search for best-fitting values and confidence ranges by calculating $\chi^2$ using parameter grids spanning $9.8\leq\log M_{\rm min} [h^{-1}M_\odot]\leq 13.6$ and $-0.6\leq \alpha_{\rm s}\leq 1.29$. The upper bounds for both parameters are chosen such that the parameter region accommodates $\Delta \chi^2=4.6$ for all subsamples. The lower bound of $\log M_{\rm min}$ is imposed as below this minimum mass the analytical $b(M_{\rm h})$ relation flattens or even reverses and is not well calibrated. Below the imposed lower bound of $\alpha_{\rm s}$ the resulting CCF becomes insensitive to the value of $\alpha_{\rm s}$.

Note that there are 113 objects in common between the 2MASS galaxy sample and \textit{Swift}/\textit{INTEGRAL} AGN sample.
In principle, this affects the satellite--satellite pair counts in the one-halo term calculation
of the CCF in the HOD modelling (see paper II). However, since only 0.3\% of the 2MASS galaxies are
common, the effect of these objects in the pair count is negligible.

\begin{figure}
\resizebox{\hsize}{!}{ \includegraphics[bbllx=69,bblly=210,bburx=514,bbury=645,angle=0]{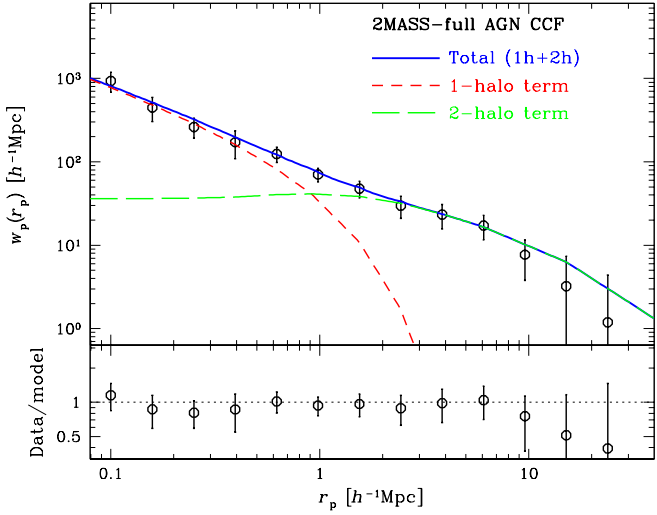}}
\caption{The measured CCF between 2MASS galaxies and the full AGN sample, shown with the best-fitting HOD model and fit residuals displayed as data/model. The symbols and line styles have the same meanings as in Fig.~\ref{fig:2massacfhod}.}
\label{fig:agnall_ccfhod}
\end{figure}

\begin{figure*}
\resizebox{\hsize}{!}{ \includegraphics[bbllx=34,bblly=292,bburx=562,bbury=569,angle=0]{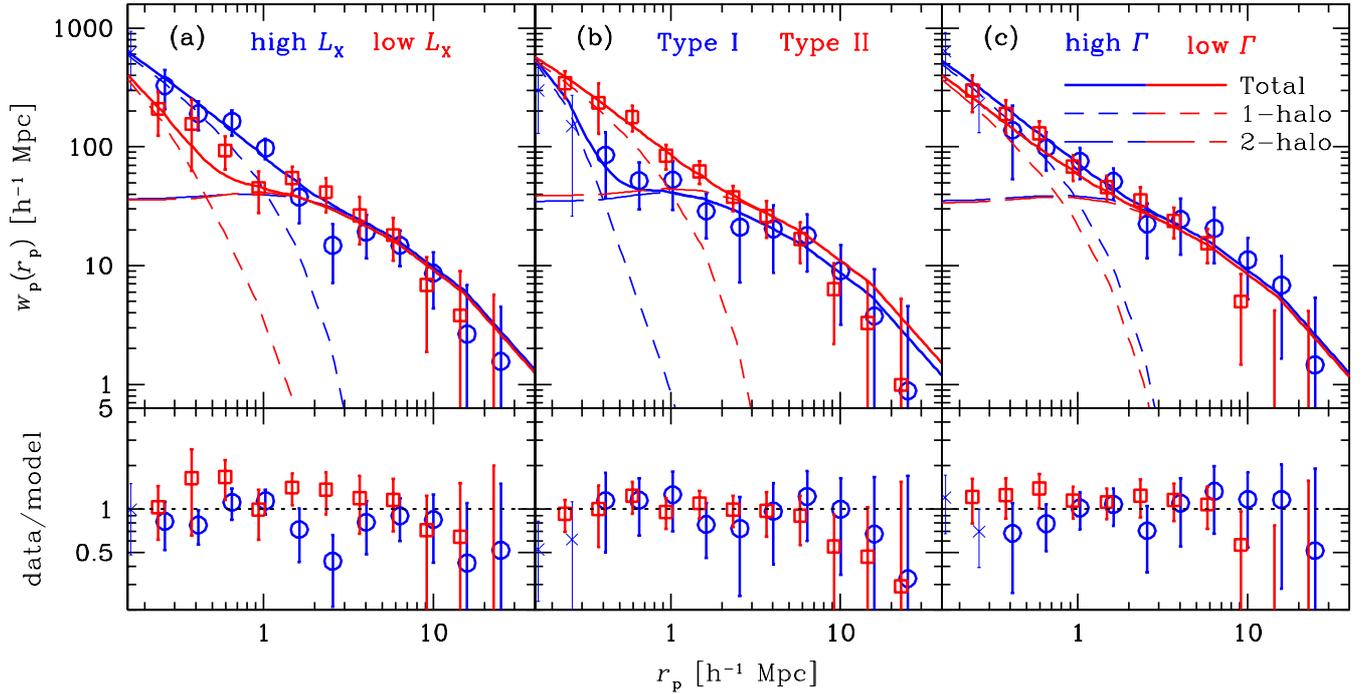}}
\caption{Similar to Fig.~\ref{fig:agnall_ccfhod}, here showing a comparison of the best-fitting HOD models and fit residuals for the $L_{\rm X}$ samples (panel a), type I versus type II samples (panel b) and X-ray photon index $\Gamma$ (panel c). Data points shown as crosses are excluded in the $\chi^2$ fitting process as they represent bins with less than 16 pairs.}
\label{fig:subsamples_ccfhod}
\end{figure*}

\section{Results}
\label{results}
\begin{table*}
	\centering
	\caption{Properties of the AGN samples and their derived HOD modelling quantities.}
	\label{overview}
	\begin{tabular}{lccccccc} 
		\hline
AGN sample & Number      & Mean         &  Median log    & Median & $b(z)$ &  $\log\,M_{\rm h}^{\rm typ}$ & $\log\,\langle M_{\rm h}\rangle$\\
name       & of objects  &$z_{\rm eff}$   &  $L_{\rm 14-195\,keV^{\dagger}}$ & $\Gamma$ & (HOD) & ($h^{-1}$ $M_{\odot}$) & ($h^{-1}$ $M_{\odot}$)\\ \hline
Full AGN sample	& 274 & 0.028 & 43.42 & 1.92 & 1.04$^{+0.10}_{-0.11}$ & 12.84$^{+0.22}_{-0.30}$ & 13.44$^{+0.05}_{-0.07}$\\
	\\	
Low-$L_{\rm X}$ AGN sample & 139 & 0.023 & 43.17 & 1.92 & 0.98$^{+0.05}_{-0.07}$ & 12.70$^{+0.12}_{-0.22}$ & 12.94$^{+0.17}_{-0.26}$\\ 
High-$L_{\rm X}$ AGN sample$^{\ddagger}$   & 135& 0.030 & 43.69 & 1.90 & 1.04$^{+0.05}_{-0.04}$ & 12.84$^{+0.12}_{-0.10}$ & 13.55$^{+0.06}_{-0.05}$\\
\\
Type I AGN sample & 113 & 0.025 & 43.45 & 2.01 & 0.96$^{+0.05}_{-0.12}$ & 12.64$^{+0.14}_{-0.44}$ & 12.75$^{+0.18}_{-0.27}$\\ 
Type II AGN sample$^{\ddagger}$ & 157 & 0.027 & 43.36 & 1.85 & 1.17$^{+0.08}_{-0.13}$ & 13.10$^{+0.14}_{-0.26}$ & 13.56$^{+0.05}_{-0.09}$\\
\\
Low-$\Gamma$ AGN sample       & 137 & 0.028 & 43.40 & 1.76 & 0.90$^{+0.15}_{-0.05}$ &12.45$^{+0.41}_{-0.19}$ & 13.27$^{+0.07}_{-0.10}$\\ 
High-$\Gamma$ AGN sample      & 124 & 0.029 & 43.42 & 2.07 & 0.98$^{+0.15}_{-0.05}$ &12.70$^{+0.34}_{-0.16}$ & 13.45$^{+0.07}_{-0.08}$\\
		\hline
\end{tabular}
\\
\begin{flushleft}
$^{\dagger}${\footnotesize $L_{\rm 14-195\,keV}$ is given in units of erg s$^{-1}$.} \\
$^{\ddagger}${\footnotesize For the high-$L_{\rm X}$ AGN sample the best fit is pegged at the lower bound of the parameter grid, log ($M_{\rm min}\,[h^{-1}{\rm M_\odot}])=9.8$). The type I AGN sample is pegged at $\alpha_{\rm s}=-0.6$.} \\ 
\end{flushleft}
\end{table*}

Fig.~\ref{fig:2massacfhod} shows our measured 2MASS galaxy ACF with the 
best-fitting HOD model and fit residuals. The best-fit HOD model (with $\chi^2/{\rm d.o.f.}=0.99$) 
corresponds to log $M_{\rm min}=\log M_0=12.50^{+0.23}_{-0.01}$, $\sigma_{\log M}=0.42^{+0.31}_{-0.01}$, log $(M_1^\prime/M_0)=1.10^{+0.01}_{-0.27}$ and $\alpha_{\rm s}=1.13^{+0.05}_{-0.07}$. The quoted uncertainty intervals correspond to $\Delta \chi^2<7.78$ (90\% confidence region in four free parameters). 
The large-scale bias parameter of our 2MASS galaxy sample is $b=1.25^{+0.02}_{-0.03}$.

The magnitude limit of our 2MASS galaxy sample is $M_{\rm K_s}=-24.4$.
We use the stellar population model of \cite{worthey_1994}\footnote{\tt{http://astro.wsu.edu/dial/dial\_a\_model.html}} to determine a rough estimate on the corresponding SDSS $r$-band magnitude. We use an old stellar population model (10 Gyr, solar metallicity), which leads to $r-K=3$ and predicts that our 2MASS galaxies are brighter than $M_{\rm r} \sim -21.4$. Indeed, our HOD results for 2MASS galaxies agree well with the HOD results presented in \cite{zehavi_zheng_2011} (see their table~3) for $M^{\rm max}_{\rm r} = -21$. Recall that we apply an upper cut on $M_{\rm K_s} = -25.9$, while the luminosity-threshold samples in \cite{zehavi_zheng_2011} have only a lower $M_{\rm r}$ limit. Thus, our 2MASS sample excludes the brightest and most massive galaxies. This naturally drives the HOD results to lower values in log $M_{\rm min}$ than the $M^{\rm max}_{\rm r} = -21.5$ sample in \cite{zehavi_zheng_2011}.

The best-fitting HOD model to the CCF between 2MASS galaxies and the full AGN sample is shown in 
Fig.~\ref{fig:agnall_ccfhod}. From this fit we derive the large-scale bias parameter listed in Table~\ref{overview} and confidence contours for both the minimum DMH mass needed to harbour an AGN and the satellite slope $\alpha_{\rm s}$ (see Fig.~\ref{fig:contour_plots} panel a). 
The measured CCF and the best-fitting HOD model for the subsamples defined with respect to 
$L_{\rm X}$, AGN type classification and X-ray photon index $\Gamma$ are presented in Fig.~\ref{fig:subsamples_ccfhod}. The HOD model parameter confidence contours are shown in  
Fig.~\ref{fig:contour_plots}. As discussed below, for the subsamples defined by $L_{\rm X}$ and AGN type classification, these two-dimensional contours have the advantage of showing a clear difference between the subsamples which is not detected when collapsing all information into a one-dimensional parameter such as the large-scale bias.

For each sample we list in Table~\ref{overview} the number of objects, the mean effective redshift of $N_{\rm CCF}(z)$, the median 14--195 keV luminosity, the median photon index $\Gamma$ and the HOD-derived large-scale bias parameter. Since we measure the CCF, the resulting effective redshift distribution for the clustering signal is determined by both the redshift distribution of the tracer set and the AGN sample: $N_{\rm CCF}(z) = N_{\rm tracer}(z)*N_{\rm AGN}(z)$.

From the HOD model, the bias parameter can be calculated using 
\begin{equation}
b_{\rm HOD}=\frac{\int b(M_{\rm h}) \langle N_{\rm A}\rangle (M_{\rm h}) \phi(M_{\rm h})dM_{\rm h}}
                      {\int \langle N_{\rm A}\rangle (M_{\rm h}) \phi(M_{\rm h})dM_{\rm h}},
\label{eq:hodbias}
\end{equation}
where $\phi(M_{\rm h})$ is the halo mass function. Defining 
$b_{\rm OBS,HOD}$ as the bias parameter obtained from the best-fitting HOD model to the observed CCF using equation~\ref{eq:hodbias}, we then define the typical DMH mass $M_{\rm h}^{\rm typ}$ by $b(M_{\rm h}^{\rm typ})=b_{\rm OBS,HOD}$. 
This conversion is performed using the same halo mass function, linear power-spectrum and $b(M)$ relation as those used in the HOD calculation. For all AGN samples, the HOD calculations are based on $z=0.02$. The values given for $M_{\rm h}^{\rm typ}$ increase by only 0.01 at $z=0.03$. 

Note that the conversion between the measured $b_{\rm OBS,HOD}$ to $M_{\rm h}^{\rm typ}$ in our previous papers was 
based on the analytical approximation of the $\nu-M_{\rm h}$ relation by \citet{vandenbosch_2002}, where $\nu$ is 
the ratio of the spherical overdensity required for collapse to the root mean square 
density fluctuation on the scale $r$ of the initial size of the object $M_{\rm h}$.  However, in this work 
this conversion is based on the linear power spectrum generated by the software {\tt camb} (see above), 
to be consistent with the HOD model calculation itself.
This change in procedure results in approximately 0.2 dex higher $M_{\rm h}^{\rm typ}$ for the same bias value; for example, the full AGN sample would have log ($M_{\rm h}^{\rm typ} [h^{-1} M_{\odot}])= 13.02$. This is important to consider when comparing $M_{\rm h}^{\rm typ}$ from our various papers.

We also calculate the mean DMH mass from the HOD model, using 
\begin{equation}
\langle M_{\rm h}\rangle=\frac{\int M_{\rm h} \langle N_{\rm A}\rangle (M_{\rm h}) \phi(M_{\rm h})dM_{\rm h}}
                      {\int \langle N_{\rm A}\rangle (M_{\rm h}) \phi(M_{\rm h})dM_{\rm h}}.
\label{eq:meanmass}
\end{equation}
 The mean halo mass is similar to $M_{\rm h}^{\rm typ}$ in that they are both representative halo masses of the sample. 
Comparing equations~\ref{eq:hodbias} and \ref{eq:meanmass}, one can see that the mean mass gives 
more weight to higher DMH masses, as, for example, $b(M_{\rm h})$ increases only from $\sim 0.8$ to $\sim 1.1$ as $\log \langle M_{\rm h}\,[h^{-1}M_\odot] \rangle$ increases from 12.0 to 13.0.  
We note that $M_{\rm h}^{\rm typ}$ is almost solely determined by the two-halo term and is thus insensitive to the CCF in the one-halo regime, while both the one-halo and two-halo terms affect $\langle M_{\rm h}\rangle$.

For the full AGN sample, we find a large-scale bias parameter of 1.04$^{+0.10}_{-0.11}$.
Comparing only the bias parameters, we find that none of the AGN subsamples defined by $L_{\rm X}$, type or $\Gamma$ show statistically significant differences in the large-scale bias. All differences are $<$1.5$\sigma$ when considering the combined uncertainties. 
However, comparing the subsamples defined by $L_{\rm X}$ and type, there are significant differences 
in $\log \langle M_{\rm h} \rangle$.
Type I and type II AGN differ in their mean halo mass $\langle M_{\rm h}\rangle$ at the  4.0$\sigma$ level in that 
type II AGN have a larger $\langle M_{\rm h}\rangle$ than type I AGN. Low-$L_{\rm X}$ AGN also have a lower 
$\langle M_{\rm h}\rangle$ than high-$L_{\rm X}$ AGN with a 3.4$\sigma$ difference. 

Figs~\ref{fig:subsamples_ccfhod}(a) and (b) and \ref{fig:contour_plots}(a) and (b) show that these significant 
differences in the clustering properties are caused by differences in the one-halo term, i.e., the CCF contributed 
by pairs within the same DMH. Low- and high-$L_{\rm X}$ AGN populate DMHs differently, in that higher {\it minimum} 
DMH masses are required to host low-$L_{\rm X}$ AGN compared to their high-$L_{\rm X}$ counterparts. 
In addition, with increasing DMH mass the fraction of high- to low-$L_{\rm X}$ AGN in satellite galaxies increases, 
as seen in the different values of $\alpha_{\rm s}$. 
Type I and type II AGN require similar $M_{\rm min}$ to host such AGN (see Fig.~\ref{fig:contour_plots}). 
However, they occupy DMHs differently, in that with increasing DMH mass the number of type I AGN in satellite galaxies 
decreases while the number of type II AGN rises proportionally with increasing DMH mass.

\begin{figure*}
\includegraphics[width=0.9\textwidth,angle=0]{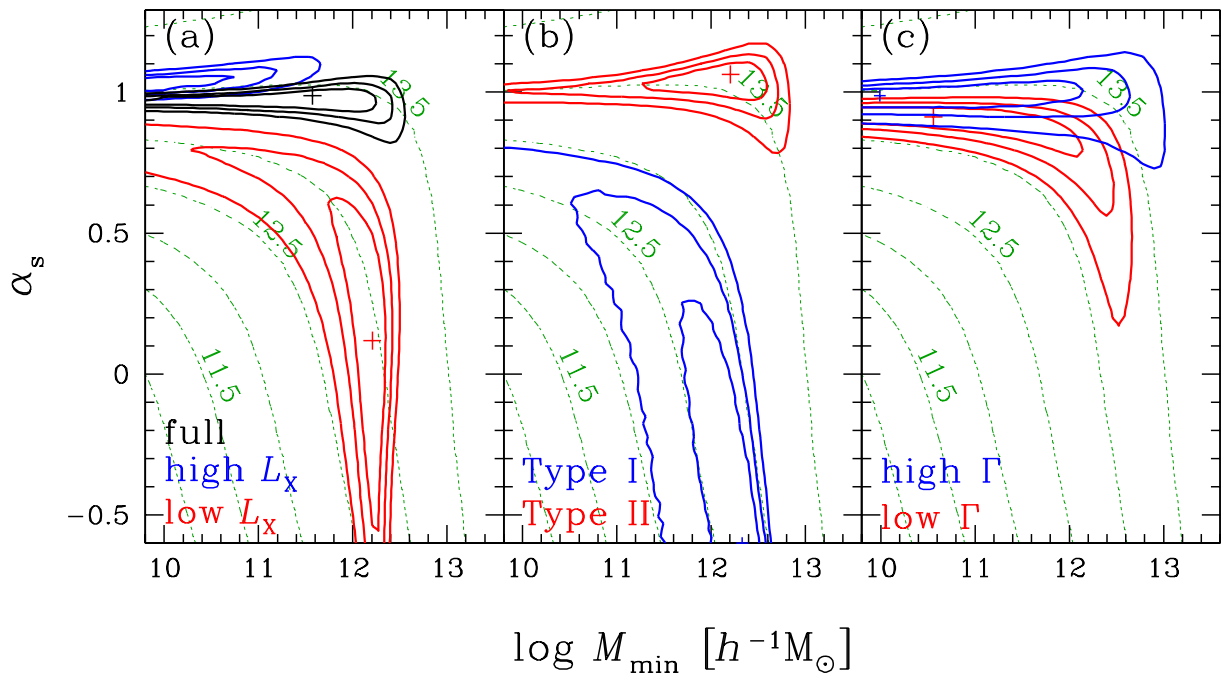}
\caption{Confidence contours of the HOD-derived parameters $\alpha_{\rm s}$ (satellite slope) and log $M_{\rm min}$ (minimum halo mass to host an AGN) for the (a) full (black), low and high $L_{\rm X}$, (b) type I and type  II classified and (c) X-ray photon index divided samples. The colour coding of the subsamples is identical to the ones used in Fig.~\ref{fig:subsamples_ccfhod}.
The confidence intervals correspond to $\Delta \chi^2=$1.0, 2.3 and 4.6 levels. 
The green dashed lines show the mean DMH mass in $\log \langle M_{\rm h} \rangle [h^{-1} M_{\odot}]$ derived from the model parameters as labelled.
}
\label{fig:contour_plots}
\end{figure*}


\subsection{Robustness of Clustering Measurements}
\label{compare_gal_lowhighz}
\label{robust}

We have paid considerable attention in this paper to creating robust clustering measurements, including careful sample selection 
and testing our results for possible systematic errors.  
First, as discussed above, we chose a redshift and absolute luminosity range for the 2MASS galaxies such that splitting the full sample into lower and higher redshift bins results in only marginal differences in the clustering signal. For the lower redshift 2MASS sample we use a redshift range of $0.007<z\le 0.025$ and for the higher redshift sample we use $0.025 <z<0.037$. This results in 2681 and 6470 galaxies in the lower and higher redshift samples, respectively. Considering the available comoving volumes within these redshift intervals, the two projected correlation functions have very similar S/Ns, and all data points for the lower and higher redshift 2MASS correlation functions agree well within their 1$\sigma$ uncertainties.  

Secondly, we tested removing all \textit{INTEGRAL} AGN from the full AGN sample and subsamples. The resulting bias values agree with those listed in Table~\ref{overview} within their combined 1$\sigma$ uncertainties. For example, the type I and type II AGN subsample using only the 70-month \textit{SWIFT}/BAT catalogue have bias values of $b_{\rm typeI}=0.98^{+0.05}_{-0.11}$ and $b_{\rm typeII}=1.19^{+0.11}_{-0.07}$. 
The corresponding mean halo masses for type I and II AGN are $\langle M_{\rm h, typeI}\rangle = 12.81^{+0.13}_{-0.27}$ and $\langle M_{\rm h, typeII}\rangle = 13.60^{+0.08}_{-0.06}$. The similarity of these values to the ones listed in Table~\ref{overview} indicates that our results are robust to the inclusion of \textit{INTEGRAL} AGN.  

Thirdly, to further check how sensitive the best fit and corresponding contours are to moderate changes in the sample 
selection, we randomly removed 15\% of objects in each of the type I and type II subsamples.
We then recomputed the CCF and reran the HOD modelling and 
find bias values of 0.93$^{+0.08}_{-0.09}$ for type I AGN and 1.06$^{+0.14}_{-0.06}$ for type II AGN (which differ from each other at the 1$\sigma$ level).  The resulting mean halo masses of $\langle M_{\rm h, typeI}\rangle = 12.72^{+0.22}_{-0.30}$ and $\langle M_{\rm h, typeII}\rangle = 13.50^{+0.09}_{-0.03}$ differ at a 3.5$\sigma$ level and the HOD-derived contours look almost identical to the ones shown in Fig.~\ref{fig:contour_plots} (panel b). We also reran the HOD modelling using uncorrelated $\chi^2$ fitting and confirm all of our results within 1$\sigma$ of the bias values and mean halo masses listed in Table~\ref{overview} and recover similar confidence contours.

Fourthly, we tested excluding in the HOD fitting the scales which could be affected by the halo--halo collision. Again, the resulting bias values and confidence contours agree well within their 1$\sigma$ uncertainties.  

Fifthly, we check whether or not the difference in redshift and X-ray luminosity between type I and II AGN (see Figs~\ref{AGN_redshift} and \ref{AGN_LX}) causes the discrepancy in the clustering properties. For this test, we create matched redshift distributions for type I and II AGN. This matching of the subsamples is a commonly used method in clustering measurements to remove possible observation biases or correlations between parameters (e.g., \citealt{coil_georgakakis_2009}; \citealt{krumpe_miyaji_2015}). 
In the case of redshift matching, we trim the type II AGN 
redshift range to that of the type I AGN.  Then, if there are more type II AGN in a given redshift bin than type I AGN, we randomly select the same number of type II AGN as type I AGN.  If there are fewer type II AGN than type I AGN, we upweight type I AGN to have the same number as type II AGN.  This ensures that the type I and II AGN have the same redshift distribution in our clustering measurements.
We also test randomly selecting a subset of AGN of a given type so as to the match the number of AGN of the other type, in a given redshift bin. The identical procedures are applied for matching the luminosity distributions. 

In the cases of matched redshift distributions for type I and type II AGN, we do not detect any significant deviation from the full type I and type II samples in Table~\ref{overview}. All realizations of type I samples (based on matched distributions) have  
$\log\,\langle M_{\rm h}\rangle = 12.70-12.81$ $h^{-1}$ $M_{\odot}$ with maximal uncertainties of $^{+0.27}_{-0.55}$. The corresponding type II samples result in $\log\,\langle M_{\rm h}\rangle = 13.45-13.56$ $h^{-1}$ $M_{\odot}$ with maximal uncertainties of $^{+0.11}_{-0.13}$.
In particular, the constraints on $\alpha_{\rm s}$ for type I and type II are very similar to those shown in Fig.~\ref{fig:contour_plots}.

Fig.~\ref{AGN_LX} shows that at extremely low and high X-ray luminosities there is little to no overlap between type I and type II AGN. Thus, we restrict the X-ray luminosity range used for matching to $42.6 \le$ log $(L_{\rm 14-195\,keV}/$[erg s$^{-1}$]) $\le 44.6$. This restriction and the matching procedure itself leads to a significant loss in the number of sources in the type I and II samples. Consequently, the error bars on our individual measurements increase substantially. 
For the same reason, the confidence contours as shown in Fig.~\ref{fig:contour_plots} (panel b) increase but the result that type I and type II AGN have significantly different clustering remains.

Sixthly, we also test whether the difference in the HOD parameters between the low- and high-$L_{\rm X}$ samples is due to differences in their redshift distributions. The matching procedure described above is not useful here, as the redshift distributions between the low and high $L_{\rm X}$ samples differ substantially, and more than half of the AGN would have to be removed to match these distributions. Thus, we create three redshift subsamples as follows: (i) $0.007 \leq z < 0.017$, (ii) $0.017 \leq z < 0.027$, and (iii) $0.027 \leq z \leq 0.037$. In each redshift range, we determine the median $L_{\rm X}$ and divide the sample into low- and 
high-$L_{\rm X}$ samples. Finally, we add all low- and high-$L_{\rm X}$ subsamples of the three individual redshift ranges. This approach produces very similar redshift distributions between the final low- and high-$L_{\rm X}$ samples. 

For these samples, we find $\log\,\langle M_{\rm h,low L_{\rm X}}\rangle = 12.95^{+0.21}_{-0.47}$ $h^{-1}$ $M_{\odot}$ and $\log\,\langle M_{\rm h,high L_{\rm X}}\rangle = 13.61^{+0.04}_{-0.07}$ $h^{-1}$ $M_{\odot}$. This is in excellent agreement with the values given in Table~\ref{results}. The resulting HOD contours are very similar to the ones shown in Fig.~\ref{fig:contour_plots} (panel a). Only the $\alpha_{\rm s}$ best-fitting value of the low-$L_{\rm X}$ sample shifts to $\alpha_{\rm s} \sim 0.68$.

In summary, we conclude that our results are not significantly impacted by systematic effects and are robust to moderate changes in our methodology and sample selection. In particular, the significant clustering difference in the mean halo masses of type I and type II AGN remains when modifying the sample selection.  


\section{Discussion}

For the full hard X-ray-selected AGN sample we find slightly lower bias values than those reported in 
\cite{cappelluti_ajello_2010}. However, the results differ by only 1.4$\sigma$, and this small difference is not surprising 
as different methods are used in each study to compute the bias. \cite{cappelluti_ajello_2010} performed power-law fits to the projected ACF, while we use the HOD modelling approach. 
For this comparison, we use bias values rather than estimated $M_{\rm h}^{\rm typ}$, as different methods are used in the literature to compute  $M_{\rm h}^{\rm typ}$, which can lead to severe systematics and misinterpretations (see also the discussion on this issue in paper IV, end of Section~5).

Although these differences in the methodology used to convert bias into $M_{\rm h}^{\rm typ}$ exist in the literature, 
here we can confirm the overall picture of AGN clustering that has emerged, in that even in the very local Universe 
AGN prefer DMH masses of $M_{\rm h}^{\rm typ}\sim 10^{13}$ $h^{-1}$ $M_{\sun}$, as they do at higher redshifts. 
In Fig.~\ref{fig:b_z} we show local and low-redshift measurements of X-ray-selected AGN that are all based on the 
cross-correlation technique. The finding that AGN inhabit group-mass haloes holds at least up to a redshift 
of $z\sim 3$ (see, e.g., fig.~5 of \citealt{allevato_civano_2016}). When comparing the clustering measurements of AGN to those of blue and red galaxies (e.g., \citealt{madgwick_hawkins_2003}, \citealt{zehavi_zheng_2005}, \citealt{hickox_jones_2009}), local AGN seem to be hosted by a mix of blue and red galaxies. This finding is consistent with that of \cite{li_kauffmann_2006}, who find that narrow-line AGN host galaxies and inactive control galaxies populate DMH of similar mass.

\subsection{Clustering dependence on X-ray luminosity}

\begin{figure}
\resizebox{\hsize}{!}{
 \includegraphics[bbllx=56,bblly=361,bburx=541,bbury=703,angle=0]{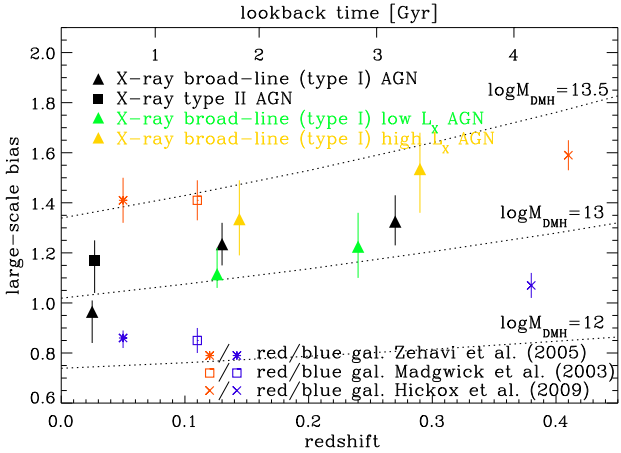}}
\caption{Large-scale bias as a function of redshift for various X-ray-selected AGN samples, compared to red and blue galaxies (red and blue symbols). Black filled triangles show X-ray-selected type I AGN from this study and \citet{krumpe_miyaji_2012,krumpe_miyaji_2015} at higher redshift. Luminosity-dependent subsamples of X-ray-selected type I AGN are shown as green (low-$L_{\rm X}$) and yellow (high-$L_{\rm X}$) filled triangles.
The black filled box shows type II AGN from this study. The biases for all AGN samples are computed by HOD modelling, while the galaxy clustering bias is based on power-law fits to the observed correlation function. The dotted lines show the expected $b(z)$ of DMH masses $M_{\rm h}^{\rm typ}$ based on \citet{sheth_mo_2001} and the improved fit for this equation given by \citet{tinker_weinberg_2005}. The masses are given in log $M_{\rm h}^{\rm typ}$ in units of $h^{-1}$ $M_{\sun}$.}
 \label{fig:b_z}
\end{figure}

Using the measured ACF of 199 \textit{SWIFT}/BAT AGN, \cite{cappelluti_ajello_2010} find an X-ray luminosity 
dependence on the clustering strength at the 1.6$\sigma$ level, using derived large-scale bias values. Here our 
bias values for the low- and high-$L_{\rm X}$ subsamples are also not statistically significantly different, with a
difference of only 0.9$\sigma$ (see Table~\ref{overview}).

At other redshifts, we detected a weak X-ray luminosity dependence (see Fig.~\ref{fig:b_z}; papers III and IV) based 
on the derived bias values. 
In \cite{krumpe_miyaji_2015} (paper IV), we showed at $z=0.16-0.36$ that the physical origin of high-$L_{\rm X}$ AGN being more clustered than low-$L_{\rm X}$ AGN is a difference in black hole mass, i.e., larger supermassive black holes reside in larger DMH masses. It is therefore not that an overdensity of galaxies leads to a higher accretion rate on to the supermassive black hole(s). 

The low number of objects in the low- and high-$L_{\rm X}$ subsamples in this paper might hamper our ability to 
detect any luminosity-dependence in the clustering strength in the local Universe.  We will need to rely on much 
larger AGN samples available in the future to test this.  For comparison, 1552 RASS AGN and $\sim$46\,000 
luminous red galaxies were used at $z=0.16-0.36$ to detect the weak X-ray luminosity dependence that was observed.

However, Fig.~\ref{fig:subsamples_ccfhod} (a) shows differences between the $L_{\rm X}$ subsamples in the one-halo term regime, and the 
$\log \langle M_{\rm h} \rangle$ value of the high-$L_{\rm X}$ sample is larger than that of the low-$L_{\rm X}$ 
sample by $\sim 3.4\sigma$. 
The confidence contours shown in Fig.~\ref{fig:contour_plots} for these two subsamples occupy completely separate regions of this space. In particular,
$\alpha_{\rm s}\sim 1$ for the high-$L_{\rm X}$ subsample, which is higher than that for the low-$L_{\rm X}$ sample, indicating a greater abundance of high-$L_{\rm X}$ AGN in satellite galaxies in the most massive haloes. This contributes to the increased clustering seen in the CCF of the high-$L_{\rm X}$ sample at $r_{\rm p}\sim 1\,\, h^{-1}\, {\rm Mpc}$.
Therefore we do find significantly different clustering properties for the high- and low-$L_{\rm X}$ samples, even if this is 
not reflected in the large-scale bias term alone.

We emphasize that our other papers (papers I--IV) are all based on soft X-ray selected, optically confirmed, broad emission-line AGN. The AGN sample presented here is hard X-ray selected and a mix of $\sim$42\% type I (broad line) and $\sim$58\% type II (narrow-line) AGN. Unfortunately, the low number of objects in each type does not allow us to further divide these samples with respect to X-ray luminosity.

\subsection{Clustering dependence on black hole mass}

Using the black hole mass estimates of \cite{koss_trakhtenbrot_2017}, we test for a clustering dependence on black hole mass for local AGN. 
No significant difference is found between the bias values or $\log \langle M_{\rm h} \rangle$ 
of the low and high $M_{\rm BH}$ subsamples 
($b_{\rm low M_{BH}} = 1.03^{+0.12}_{-0.12}$, 
$b_{\rm high M_{BH}}=0.98^{+0.04}_{-0.02}$,
$\log \langle M_{\rm h,low M_{BH}} \rangle = 13.15^{+0.17}_{-0.38}$ and 
$\log \langle M_{\rm h,high M_{BH}} \rangle = 13.45^{+0.05}_{-0.02}$).

We suffer from three challenges in possibly detecting a dependence of clustering on $M_{\rm BH}$.  
First, we suffer from a further decrease in sample size as only $\sim$60\% of local \textit{SWIFT}/BAT AGN have black hole mass estimates thus far.  
Secondly, paper IV shows that the strength of the observed $M_{\rm BH}$ dependence at higher redshift 
is weak and of similar amplitude as the weak $L_{\rm X}$ dependence. As discussed above, 
we do not detect a statistically significant $L_{\rm X}$ dependence on the clustering for local AGN using the large-scale 
bias values, though we do detect differences in the one-halo term.  

Thirdly, there is a substantial difference in the mean black hole mass estimated at a given $L_{\rm X}$ for the two 
methods used in \cite{koss_trakhtenbrot_2017} (see Fig.~\ref{Mbh_LX_koss}). 
The linear fits of the black hole masses derived from the different methods each show a correlation between $L_{\rm X}$ and $M_{\rm BH}$, offset by about an order of magnitude, depending on which method is used. 
Given the large possible systematic difference in the masses derived using these two methods, we attempted to measure
the clustering dependence on $M_{\rm BH}$ using masses derived just with the stellar velocity dispersion method, which 
has more estimated black hole masses.  However, this results in too few objects to derived a constrained CCF measurement. 

We conclude that the non-existence of a clustering dependence with black hole mass in the local Universe cannot be ruled out with the present limited data.  This will be an interesting test for future studies of local AGN with larger samples with black hole mass estimates.

\subsection{Clustering differences for type I and II AGN}

In contrast to \cite{cappelluti_ajello_2010}, we do not find that type I AGN have a higher bias value than type II AGN. In detail, the bias values for type I AGN differ significantly between these two studies (by $\sim$7$\sigma$), while the values for type II AGN agree well within their uncertainties. In our directly measured CCFs between the type I and type II AGN and the 2MASS galaxy sample, the difference is clearly visible at small and intermediate scales (Fig.~\ref{fig:subsamples_ccfhod}). 
The HOD fits provide further insights into the different clustering properties of type I and type II AGN. The confidence contours of log $M_{\rm min}$ and $\alpha$ shown in Fig.~\ref{fig:contour_plots} demonstrate a clear difference in the clustering properties of type I versus type II AGN, and we detect a difference in $\log \langle M_{\rm h} \rangle$ at the 4$\sigma$ level, similar to the difference seen between the low and high $L_{\rm X}$ samples.

Type I AGN have a much shallower satellite slope $\alpha_{\rm s}$ than type II AGN. The slope found here for type I AGN is consistent with that of type I AGN at higher redshifts (see paper II and e.g. \citealt{allevato_finoguenov_2012}). Thus, it appears that with increasing DMH mass the duty cycle of type I AGN  in satellite galaxies decreases. This could possibly be explained by AGN triggering of major and minor mergers of sub-haloes inside group mass host DMHs, as the  merging cross-section decreases in the high relative velocity encounters that are more common as the DMH mass increases \citep{altamirano2016}. At least for local type II AGN this appears not to be the dominant triggering mechanism, as they have an HOD satellite slope that is similar to that of galaxies. Thus, the type II AGN duty cycle in satellite galaxies is close to constant with increasing DMH mass. Therefore, type II AGN should have a higher duty cycle in more massive DMHs than type I AGN. This suggests that the dominant triggering mechanism of type II AGN in groups and clusters is not mergers but more likely disturbances due to close encounters and/or secular processes. 

The type I and type II AGN samples have very similar median X-ray luminosities (see Table~\ref{overview}) as well as similar X-ray luminosity distributions (Fig.~\ref{AGN_LX}). We tested matching their redshift and X-ray luminosity distributions and find similar results (see Section~\ref{robust}). Thus, the different clustering properties of type II AGN compared to type I AGN cannot be explained by the observed X-ray luminosity dependence of the clustering signal. 

According to a simple orientation-based AGN unification scheme, in
which type I and type II AGN should have similar distributions 
of properties such as redshift, luminosity and black hole mass, no statistical 
difference is expected in their clustering behaviour. 
Thus, our results contradict a simple AGN unification model, at least for the local Universe. 

Other studies also find evidence that type I and type II AGN might reside in different environments. \cite{villarroel_korn_2014} conduct number counts around type I (broad-line) and type II (narrow-line) AGN at $0.03 < z <0.2$. Their results are also inconsistent with a simple orientation-based AGN unification scheme. 
They find, for example, that the average colour of the neighbours is redder around type I AGN than around type II and that the fraction of AGN in spiral host galaxies depends on AGN type. Other studies use different samples but similar methods and find contradictory results. For example, \cite{jiang_wang_2016} find that type II AGN have significantly more satellites than type I AGN do, while \cite{gordon_owers_2017} cannot confirm such a difference. 

As shown in Fig.~\ref{NH_type}, separating AGN by type is similar to separating AGN by the amount of absorbing column density $N_{\rm H}$.  This suggests that future studies may be able to detect a weak clustering dependence on $N_{\rm H}$ in the local Universe, as obscured AGN should show similar clustering properties as type II AGN.

\subsection{Clustering Dependence on Photon Index}

\cite{trakhtenbrot_ricci_2017} analysed 228 BAT AGN in the redshift range of $0.01<z<0.5$ for which black hole mass estimates are available. They found a weak correlation between photon index and accretion rate of the black holes in that, on average, higher photon index corresponds to higher accretion ratio relative to Eddington. Thus, measuring the clustering dependence on the photon index might reflect the dependence on accretion rate as well. We note that \cite{trakhtenbrot_ricci_2017} determine their photon index in the 2--10 keV band, unlike what is used in this study.

Here we do not find a significant clustering dependence on photon index determined in the 14--195 or 20--200 keV bands. Thus it would appear that there is likely no strong clustering dependence on accretion ratio relative to Eddington. In paper IV, we also tested for a clustering dependence on accretion ratio using a soft X-ray-selected broad emission-line AGN sample and did not find a correlation either.   
\cite{aird_coil_2012} show that the accretion rate relative to Eddington does not strongly depend on stellar mass of the host galaxy. Thus, it might not be surprising that the clustering dependence on accretion ratio, if it exists at all, should be very weak.


\section{Conclusions}

We explore the clustering properties of hard X-ray-selected AGN in the local Universe. 
This study is the most precise clustering measurement made to date for AGN at very low redshift; 
the median redshift of $z =0.024$ corresponds to a look-back time of only 
330 million years. Selecting AGN using hard X-rays (14--195 and 20--200 keV bands) 
allows a measurement of AGN clustering that is unbiased with respect to absorption. 
We achieve the high precision by using a cross-correlation approach, in that we compute the two-point CCF between the AGN sample and a much larger sample of 2MASS galaxies 
which acts as a tracer set of the large-scale matter distribution in the local Universe. We then perform HOD modelling to the CCF to derive the large-scale bias parameter and determine the distribution of AGN within DMHs. As a by-product of our cross-correlation technique, we also report here for the first time an HOD model of 2MASS-selected galaxies, for $0.007\le z\le 0.037$ and $-24.4 \le M_{\rm K_s} \le -25.9$ (see Section~\ref{results}).  

We analyse the clustering properties not only for the full AGN sample but also for subsamples defined by X-ray luminosity, type I versus type II classification, photon index of the X-ray spectrum and black hole mass.  
The full AGN sample ($0.007 < z < 0.037$) has a bias parameter of $b=1.04^{+0.10}_{-0.11}$, which corresponds to a typical host DMH mass of $M_{\rm h}^{\rm typ}=12.84^{+0.22}_{-0.30}\,h^{-1}\, M_{\odot}$.

Comparing the large-scale bias values, we do not find statistically significant differences in the subsamples as a function of X-ray luminosity, AGN type classification, photon index and black hole mass. In \cite{krumpe_miyaji_2015} we showed that at $z\sim 0.3$ only a weak clustering dependence as a function of X-ray luminosity and black hole mass exists. 
The moderate sample sizes used here (when splitting the full AGN sample with respect to the different AGN parameters) may hamper our ability to detect such a weak clustering dependence in the local Universe.
However, the joint two-dimensional space that is fit by our AGN HOD model (which constrains the minimum DHM mass required to host an AGN and the satellite slope), as well as the mean halo mass, do show significant clustering differences for AGN as a function of X-ray luminosity.  We find that high X-ray luminosity AGN have a larger mean halo mass than their low-luminosity counterparts at the 3.4$\sigma$ level.

We also find significant differences in the clustering properties of type I and type II AGN.  While similar 
minimum halo masses are required to host type I and type II AGN, type II AGN have a higher {\it mean}
 halo mass than type I AGN, at the 4.0$\sigma$ level. 
As found previously at other redshifts (\citealt{miyaji_krumpe_2011}; \citealt{allevato_finoguenov_2012}), our HOD model for 
local type I AGN indicates that above DMH masses of 
$M_{\rm h} \sim 10^{12}$ h$^{-1}$ $M_{\odot}$ the duty cycle of AGN in satellite galaxies decreases 
with increasing halo mass, in that the slope of the HOD is shallower for AGN than for galaxies.
In contrast, local type II AGN have an HOD that is very similar to that of galaxies. The average number of type II AGN in satellite galaxies increases approximately proportionally to DMH mass, as is observed for galaxies. Thus, the ratio of type I to II AGN should decrease with DMH mass. These differences in the clustering properties of type I and type II AGN are at odds with a simple orientation-based AGN unification scheme. 

This study demonstrates that testing the clustering properties of AGN with respect to AGN parameters such as redshift, luminosity and classification not only enhances our understanding of how AGN populate DMHs through cosmic time, but that such measurements can constrain AGN models such as the unification model which, at first glance, may seem to have no connection to large-scale structure. Extending this approach with much larger samples will allow us to significantly constrain physical models of AGN triggering and evolution.

\section*{Acknowledgements}

We would like to thank Lutz Wisotzki for helpful discussions and supporting this project. 
MK acknowledges support by DFG grant KR 3338/3-1. TM and HA thank 
for financial support by CONACyT Grant Cient\'ifica B\'asica 
\#252531, UNAM-DGAPA Grants PAPIIT IN104113 and IN104216. 
This publication makes use of data products from the 2MASS, which is a joint project of the University of Massachusetts and the
Infrared Processing and Analysis Center/California Institute of Technology,
funded by the National Aeronautics and Space Administration and the National
Science Foundation. This work is based on data taken by \textit{SWIFT}, a MIDEX mission led by NASA with participation of Italy and the UK.
We thank Angela Malizia for providing the \textit{INTEGRAL}/IBIS catalogues. 
This study is based on \textit{INTEGRAL} data. 
This research has made use of data and/or software provided by the High Energy Astrophysics Science Archive Research Center (HEASARC), which is a service of the Astrophysics Science Division at NASA/GSFC and the High Energy Astrophysics Division of the Smithsonian Astrophysical Observatory.
This research has made use of data and/or software provided by the High Energy Astrophysics Science Archive Research Center (HEASARC), which is a service of the Astrophysics Science Division at NASA/GSFC and the High Energy Astrophysics Division of the Smithsonian Astrophysical Observatory. 
We also thank the referee for a useful report.







\bsp	
\label{lastpage}
\end{document}